\documentstyle[aps,prd,amsopn]{revtex}

\begin{document}
\draft

\title{Prescriptionless light-cone integrals}
\author{A. T. Suzuki\footnote{E-mail:suzuki@ift.unesp.br} and A. G. M.
Schmidt\footnote{E-mail:schmidt@ift.unesp.br}} \address{ Instituto de
F\'{\i}sica Te\'orica, Universidade Estadual Paulista, R.Pamplona, 145 S\~ao
Paulo - SP CEP 01405-900 Brazil}\date{today}
\maketitle

\begin{abstract}
Perturbative quantum gauge field theory seen within the perspective of physical
gauge choices such as the light-cone entails the emergence of troublesome poles
of the type $(k\cdot n)^{-\alpha}$ in the Feynman
integrals, and these come from the boson field propagator, where $\alpha =
1,2,\cdots$ and $n^{\mu}$ is the external arbitrary four-vector that defines
the gauge proper. This becomes an additional hurdle to overcome in the
computation of Feynman diagrams, since any graph containing internal boson
lines will inevitably produce integrands with denominators bearing the
characteristic gauge-fixing factor. How one deals with them has been the
subject of research for over decades, and several prescriptions have been
suggested and tried in the course of time, with failures and successes. 

However, a more recent development in this front which applies the negative
dimensional technique to compute light-cone Feynman integrals shows that we can
altogether dispense with prescriptions to perform the calculations. An
additional bonus comes attached to this new technique in that not only it
renders the light-cone prescriptionless, but by the very nature of it, can also
dispense with decomposition formulas or partial fractioning tricks used in the
standard approach to separate pole products of the type $(k\cdot
n)^{-\alpha}[(k-p)\cdot n]^{-\beta}$, $(\beta = 1,2,\cdots)$.

In this work we demonstrate how all this can be done.

\end{abstract}

\vspace{.5cm}


\section{Introduction.}

Light-cone gauge for gauge field theories is probably one of the most widely
used among the algebraic non-covariant gauges. Its popularity has known ups and
downs along its history. Among the ups are that the emerging propagator has a
deceivingly simple structure compared to other non-covariant choices, the
decoupling of Fadeev-Popov ghosts from the physical fields, and the possibility
of describing and modeling complex supersymmetric string theories in it. The
ugly side of the coin is represented by the subtle $(k\cdot n)^{-\alpha}$
singularities present in all the physical amplitudes described within it. Such
complication demanded {\em ad hoc} prescriptions to handle the singularity in a
mathematically consistent way. Apart from the fact that such expedient has to
be carried out by hand, it was soon realized that it was not enough to be
mathematically well-defined, it had to be physically consistent as well. Thus,
not any prescription is suitable, but only {\it causal} prescriptions are
eligible for the light-cone gauge.

Probably the major breakthrough in recent years along this line is the
realization that $D$-dimensional Feynman integrals can be analytically
continued to negative dimensions and performed there and then brought back to
positive dimensionality \cite{halliday,flying}. Negative dimensional integration
method (NDIM) is tantamount to the performing of fermionic integration in
positive dimensions \cite{halliday2}. This, can be applied to light-cone
integrals with surprising effects. No prescription is called for the
computation \cite{probing} and moreover, as it can be shortly seen, it can
dispense altogether with the necessity of partial fractioning products of
gauge-dependent poles \cite{eprint}, a {\em condition sine qua non} when one
resorts to the use of prescriptions.

In this work we shall demonstrate the two surprising features of NDIM when
employed in the light-cone context: no prescriptions and no partial
fractionings are needed. Our lab-testing is performed taking the simplest
scalar and tensorial structures for one-loop integrals.

\section{One-loop light-cone gauge loop integrals.}

First of all, let us make things more concrete, by analysing the framework
of vector gauge fields, e.g. the pure Yang-Mills fields, where, after taking
the limit of vanishing gauge parameter, the propagator reads:

\begin{equation}
\label{prop}
D^{ab}_{\mu\nu}(k)=\frac{-i\,\delta^{ab}}{k^2+i\varepsilon}
\left[g_{\mu\nu}-\frac{n_{\mu}k_{\nu}+n_{\nu}k_{\mu}}{k\!\cdot\! n} \right ]
\end{equation}
where here $(a,b)$ are the gauge group indices, $n_{\mu}$ is the arbritary and
constant light-like four-vector which defines the gauge,
$n\!\cdot\!A^a(x)=0\,;\;\;\;n^2=0$. This propagator generates $D$-dimensional
Feynman integrals of the following generic form:

\begin{equation}
\label{struc}
I_{lc}=\int
\frac{d^Dk_i}{A(k_j,\;p_l)}\frac{f(k_j\!\cdot\!n^*,\;p_l\!\cdot\!n^*)}{h(k_j\!\cdot\!n,\;p_l\!\cdot\!n)},
\end{equation}
where $p_{l}$ labels all the external momenta, and $n^{*}_{\mu}$ is a null
four-vector, dual to $n_{\mu}$. A conspicuous feature that we need to note
first of all, is that the dual vector $n_{\mu}^{*}$, when it appears at all,
it does so {\em always } and {\em only} in the numerators of the integrands.
And herein comes the first seemingly ``misterious'' facet of light-cone gauge.
How come that from a propagator expression like (\ref{prop}), which contains no
$n^*$ factors, can arise integrals of the form (\ref{struc}), with prominently
seen $n^*$ factors? Again, this is most easily seen in the framework of 
definite external vectors $n$ and $n^*$. An alternative way of writing the
generic form of a light-cone integral is

\begin{equation}
\label{struc1}
I_{lc}^{\mu_1\cdots \mu_n}=\int
\frac{d^{D}k_{i}}{A(k_{j},\;p_{l})}\frac{g(k^{\mu_j}_{j},\;p_l^{\mu_l})}{h(k_{j}\!\cdot\! n,\;p_{l}\!\cdot\! n)},
\end{equation}
where the numerator $g(k^{\mu_j}_{j},\;p_l^{\mu_l})$ defines a tensorial
structure in the integral. For a vector, we have $k^{\mu}=(k^+,\;k^-,\;{\bf
k}^{\sc t})$, where $k^+=2^{-1/2}(k^0+k^{D-1})$ and 
$k^-=2^{-1/2}(k^0-k^{D-1})$. If we choose definite $n$ and $n^*$ such that
$n_{\mu}=(1,\;0,\;\cdots,\;1)$, and $n^*_{\mu}=(1,\;0,\;\cdots,\;-1)$, this
allows us to write $k^+\equiv k\!\cdot\!n$ and $k^-\equiv k\!\cdot\!n^*$. We
have therefore traced back the origin for the numerator factors containing
$n^*$. We would like to emphasize here that the presence of this $n^{*}$ in the
numerators of integrands has nothing whatsoever to do with some kind of
prescription input. It is rather an intrinsic feature of the general structure
of a Feynman integral in the light-cone gauge.

Of course, for practical reasons we illustrate the NDIM methodology picking up
only few of the scalar, vector and second-rank tensor one-loop integrals. So,
we shall be considering the following:

\begin{equation}  \label{primeira}
T_1(i,j,l) = \int d^D\! q\,\, {\bf N}(q)\,,
\end{equation}

\begin{equation}  \label{segunda}
T_1^\mu (i,j,l) = \int d^D\! q\,\, q^\mu \,\, {\bf N}(q)\,,
\end{equation}

\begin{equation}  \label{terceira}
T_1^{\mu\nu}(i,j,l) = \int d^D\! q\,\, q^\mu q^\nu \,\, {\bf N}(q)\,,
\end{equation}
where 
\[
{\bf N}(q)\equiv [(q-p)^{2i}] (q\!\cdot\! n)^j (q\!\cdot\! n^*)^l.
\]
and
\begin{equation}  \label{quarta}
T_2(i,j,l,m) = \int d^D\! q\,\, {\bf R}(q)\,,
\end{equation}

\begin{equation}  \label{quinta}
T_2^\mu (i,j,l,m) = \int d^D\! q\,\, q^\mu \,\,{\bf R}(q)\,, 
\end{equation}

\begin{equation}  \label{sexta}
T_2^{\mu\nu}(i,j,l,m) = \int d^D\! q\,\, q^\mu q^\nu\,\,{\bf R}(q)\,, 
\end{equation}
where 
\[
{\bf R}(q) \equiv [(q-p)^{2i}] (q\!\cdot\! n)^j [(q-p)\!\cdot
\!n]^l (q\!\cdot \!n^*)^m.
\]

In the first three type $T_1$ integrals, after they are computed in NDIM, only
the exponents $(i,j)$ will be analytically continued to allow for negative
values, since the original structure of the Feynman integral demands exponent
$l\geq 0$. Similarly, for the last three type $T_2$ integrals only the
exponents $(i,j,l)$ will be analytically continued to negative values whereas
$m\geq 0$. We strongly emphasize this point in view of the fact that we must
respect the very nature of the original structure for the light-cone integrals,
where factors of the form $(q\!\cdot\!n^*)$ {\em never} appears in
denominators.

Observe that we are not invoking any kind of prescription for the
$(q\!\cdot\!n)^j$ factors to solve the integrals in NDIM, since before analytic
continuation $j$ is strictly {\em positive} and there are no poles to
circumvent! This is the beauty and strength of NDIM! Neither are the
$(q\!\cdot\!n^*)^l$ numerator factors due to some sort of prescription input as
they are, e.g., in the Mandelstam-Leibbrandt (ML) treatment, where one makes
the substitution \cite{leib,bass,bass2,mandelstam,leib2,lee}

\begin{equation}
{\cal M} = \int \frac{d^D\! q}{(q-p)^2 (q\!\cdot\! n)} \longrightarrow
^{\!\!\!\!\!\!\!\!\!\!\!\!\!\! ML} \int \frac{d^D\! q \, (q\!\cdot\!
n^*)}{(q-p)^2 \left[(q\!\cdot\! n) (q\!\cdot\! n^*) +i\epsilon\right]}.
\end{equation}

Let us then evaluate the integrals using the NDIM approach. In fact, our first
integral $T_1$ has already been calculated with great details in our previous
paper \cite{probing} whose result is,
 
\begin{equation}
\label{t1}
T_1^{AC}(i,j,l) = \pi^{\omega}\chi^{i+\omega}\;(p\!\cdot\!
n)^j\,(p\!\cdot\!n^{*})^l\;\frac{(-i|2i+\omega)(-j|-i-\omega)}{(1+l|i+\omega)}\,,
\end{equation}
where 
\[
\chi \equiv \frac {2\:p\!\cdot\! n\; p\!\cdot\! n^{\ast}}{
n\!\cdot\! n^{\ast}},
\]
and the superscript ``$AC$'' means that the exponents $(i,j)$ were
analytically continued to allow for negative values, $\omega=D/2$ and we use
the Pochhammer symbol, 
\begin{equation}
(a|b) \equiv (a)_b = \frac{\Gamma(a+b)}{\Gamma(a)} .
\end{equation}

Observe that $l$ must take only positive values or zero since the Pochhammer
symbol containing $\Gamma(1+l)$ was not analytically continued.

Consider now the second integral, vectorial, given in (\ref{segunda}). For this
case, let,

\begin{equation}  \label{gmu}
G^\mu = \int d^D\! q\, q^\mu \,\exp{\left[-\alpha (q-p)^2-\beta (q\!\cdot\! n)
-\gamma (q\!\cdot\! n^*)\right]} .
\end{equation}

Introducing the standard trick of substituting the $q^{\mu}$ factor for a
derivative in $p_{\mu}$ \cite{bogoliubov}, we obtain,
 
\begin{eqnarray}
G^\mu &=& \left(\frac{\pi}{\alpha}\right)^\omega \frac{e^{-\alpha p^2}}{
2\alpha}\frac{\partial}{\partial p_\mu} \exp{\left[ \alpha p^2+ \frac{
\beta\gamma}{2\alpha} (n\!\cdot\! n^*) - \beta p^+-\gamma p^-\right]}  \nonumber
\\
&=& \left(p^\mu - \frac{\beta}{2\alpha} n^\mu - \frac{\gamma}{2\alpha}
n^{*\mu}\right) {\cal G}_0 ,
\end{eqnarray}
where $p^+=p\!\cdot\! n$ and $p^-=p\!\cdot\! n^*$, as usual in the light-cone
notation \cite{leib,bass}, and define 
\begin{equation}
{\cal G}_0 \equiv \left(\frac{\pi}{\alpha}\right)^\omega \exp{\left[ \frac{
\beta\gamma}{2\alpha}(n\!\cdot\! n^*) -\beta p^+-\gamma p^-\right] } .
\end{equation}

Now, Taylor expanding the exponential in equation (\ref{gmu}), 

\begin{equation}
G^\mu = \sum_{i,j,l=0}^\infty \frac{(-1)^{i+j+l} \alpha^i\beta^j\gamma^l}{
i!j!l!}\; T_1^\mu (i,j,l) ,
\end{equation}
and following the steps for NDIM calculation \cite{halliday} we get finally, 

\begin{equation}
\label{t1mu}
T_1^{\mu,\:AC} (i,j,l) = V_1^{\mu}\:T_1^{AC}(i,j,l) ,
\end{equation}
where 

\begin{equation}
V_1^{\mu} \equiv p^\mu - \left [\frac{(i+\omega)p^-}{(1+i+l+\omega)(n\!\cdot\!
n^*)}\right ]\;n^\mu - \left [\frac{(i+\omega)p^+}{(1+i+j+\omega)(n\!\cdot\!
n^*)}\right ]\;n^{*\mu}\:.
\end{equation}

This result is in Euclidean space and valid for positive dimension
($D=2\omega>0$), {\em negative} exponents $(i,j)$ and $l\geq 0$.

The second-rank tensor integral in (\ref{terceira}) can be evaluated in a
similar way. The only thing that need to be taken into account is that now a
second derivative is called for and the calculation becomes lengthier. We only
quote the final result:

\begin{equation}
\label{t1munu}
T_1^{\mu\nu,\:AC}(i,j,l)=V_1^{\mu\nu}\:T_1^{AC}(i,j,l),
\end{equation}
where

\begin{eqnarray}
V_1^{\mu\nu}&\equiv & p^\mu p^\nu -
\left [\frac{(i+\omega)p^+p^-}{(1+i+j+\omega)(1+i+l+\omega)(n\!\cdot\!
n^*)}\right ]\;g^{\mu\nu} \nonumber \\
&-& \left [\frac{(i+\omega)p^-}{(1+i+l+\omega)(n\!\cdot\!n^*)}\right ]\;(p^\mu
n^\nu+p^\nu n^\mu)\nonumber \\
&-& \left [\frac{(i+\omega)p^+}{(1+i+j+\omega)(n\!\cdot\! n^*)}\right ]\;(p^\mu n^{*\nu}+p^\nu
n^{*\mu})\nonumber \\
&+& \left [\frac
{(i+\omega)(1+i+\omega)p^+p^-}{(1+i+j+\omega)(1+i+l+\omega)(n\!\cdot\!n^*)^2}\right
]\;(n^{\mu}n^{*\nu}+n^{\nu}n^{*\mu}) \nonumber \\
&+&
\left [\frac{(i+\omega)(1+i+\omega)(p^-)^2}{(2+i+l+\omega)(1+i+l+\omega)(n\!\cdot\!
n^*)^2}\right ]\;n^{\mu}n^{\nu} \nonumber\\
&+&\left
[\frac{(i+\omega)(1+i+\omega)(p^+)^2}{(2+i+j+\omega)(1+i+j+\omega)(n\!\cdot\!n^*)^2}\right
]\;n^{*\mu}n^{*\nu}.
\end{eqnarray}

It can be noted that for the particular case of $i=j=-1$ the pole piece for
$\omega \rightarrow 2$ only arises in the scalar integral factor
$T_1^{AC}(i,j,l)$, equation (\ref{t1}).

Now, let us consider the integrals $\{T_2\}$. These contain two scalar products
with $n_{\mu}$, but again they are harmless in NDIM approach because their
exponents, before analytic continuation, are positive. However, in the usual
positive dimensional approach, such factors can become singular and
prescriptions become a necessity. Yet prescriptions cannot handle products; one
needs to use partial fractioning first. Thus, the recourse is to use the
so-called ``decomposition formulas'' such as (see, for example,
\cite{leib,leib-nyeo})

\begin{equation}  \label{d-formula}
\frac{1}{(k\!\cdot\! n)(p-k)\!\cdot\! n} = \frac{1}{p\!\cdot\! n} \left[ \frac{1}{
(p-k)\!\cdot\! n} + \frac{1}{k\!\cdot\! n}\right] ,\:\: p\!\cdot\! n \neq 0\,,
\end{equation}

NDIM does not require any of such partial fractionings; it can handle products
at the same time. Not only that, NDIM can handle any power of these products
simultaneously, i.e., factors of the form $(k\cdot n)^{-\alpha}[(p-k)\cdot
n]^{-\beta}$, with $(\alpha,\beta = 2,3,\cdots)$ which, of course, becomes the
more strenuously difficult to handle by partial fractioning the higher the
power we have. 

To evaluate $T_2$ using NDIM, let us then consider the Gaussian-like integral,

\begin{equation}
\label{g2}
G_2 = \int d^D\! q\, \exp{\left[ -\alpha (q-p)^2-\beta q\!\cdot\! n -\gamma
(q-p)\!\cdot\! n -\delta q\!\cdot\! n^*\right]},
\end{equation}
which yields

\begin{equation}  \label{gausst2}
G_2 = \left(\frac{\pi}{\alpha}\right)^{D/2} \exp{\left( -\beta p^+ -\delta
p^-+\frac{\beta\delta}{2\alpha}n\!\cdot\! n^* + \frac{\gamma\delta}{2\alpha}
n\!\cdot\! n^*\right)}\,.
\end{equation}

On the other hand, direct Taylor expansion of (\ref{g2}) yields

\begin{equation}
G_2 = \sum_{i,j,l,m=0}^\infty (-1)^{i+j+l+m} \frac{\alpha^i \beta^j \gamma^l
\delta^m}{i!j!l!m!}\;T_2(i,j,l,m) .
\end{equation}

Comparing both expressions and solving for $T_2(i,j,l,m)$ we get a unique
solution for a system of $4\times 4$ linear algebraic equations \cite{flying},
which analytically continued to positive dimension and {\em negative} values
for $(i,j,l)$ finally gives

\begin{equation}
\label{t2} T_2^{AC}(i,j,l,m) =\pi^\omega\,
\chi^{i+\omega}\;(p^{+})^{j+l}(p^{-})^m\;\frac
{(-i|2i+l+\omega)(-j|-i-l-\omega)}{(1+m|i+\omega)}\,. 
\end{equation}

Again, superscript ``$AC$'' means $(i,j,l)$ strictly {\em negative} and $m\geq
0$.

With help of equation (\ref{gausst2}) it is easy to solve the two remaining
integrals, whose final results we quote here:

\begin{equation}
\label{t2mu}
T_2^{\mu,\,AC}(i,j,l,m) = V_2^{\mu}\:T_2^{AC}(i,j,l,m)\,,
\end{equation}
where

\begin{equation}
V_2^\mu \equiv  p^\mu -\left [
\frac{(i+\omega)p^-}{(1+i+m+\omega)(n\!\cdot\!n^*)}\right ]\;n^{\mu}-\left
[\frac {(i+l+\omega)p^+}{(1+i+j+l+\omega)(n\!\cdot\!n^*)}\right ]\;n^{*\mu}\,,
\end{equation}
and 

\begin{equation}
\label{t2munu}
T_2^{\mu \nu,\,AC}(i,j,l,m)=V_2^{\mu\nu}\:T_2^{AC}(i,j,l,m),
\end{equation}
where

\begin{eqnarray}
V_2^{\mu\nu} &\equiv & p^\mu p^\nu - \left [\frac
{(i+l+\omega)p^+p^-}{(1+i+j+l+\omega)(1+i+m+\omega)(n\!\cdot\!n^*)}\right
]\;g^{\mu\nu} \nonumber\\
&-&\left [\frac{(i+\omega)p^-}{(1+i+m+\omega)(n\!\cdot\!n^*)}\right
]\;(p^\mu n^\nu+p^\nu n^\mu)\nonumber\\
&-&\left [\frac{(i+l+\omega)p^+}{(1+i+j+l+\omega)(n\!\cdot\!n^*)}\right
]\;(p^\mu n^{*\nu}+p^\nu n^{*\mu})\nonumber\\
&+&\left [\frac{(i+l+\omega)(1+i+\omega)p^+p^-}{(1+i+m+\omega)(1+i+j+l+\omega)
(n\!\cdot\!n^*)^2}\right ]\;(n^\mu n^{*\nu}+n^\nu n^{*\mu})\nonumber\\ 
&+&\left [\frac{(i+\omega)(1+i+\omega)(p^-)^2}{(2+i+m+\omega)(1+i+m+\omega)
(n\!\cdot\!n^*)^2}\right ]\;n^\mu n^{\nu} \nonumber\\ 
&+&\left
[\frac{(i+l+\omega)(1+i+l+\omega)(p^+)^2}{(2+i+j+l+\omega)(1+i+j+l+\omega)
(n\!\cdot\!n^*)^2}\right ]\;n^{*\mu} n^{*\nu} \,.
\end{eqnarray}

Finally, before closing this section, let us analyse (\ref{quarta}) with
momentum shift $q=p-k$, so that

\begin{equation}
\label{tau2}
T_2(i,j,l,m)=(-1)^{j+l+m}\:\tau_2(i,j,l,m)\,, \qquad {\rm or} \qquad
\tau_2(i,j,l,m)=(-1)^{-j-l-m}\:T_2(i,j,l,m)\,,
\end{equation}
where

\begin{equation}
\tau_2(i,j,l,m)=\int d^D\! k\,\,k^{2i}[(k-p)\!\cdot\! n]^j (k\!\cdot
\!n)^l [(k-p)\!\cdot \!n^*]^m.
\end{equation}

Then, we can easily write down the following results

\begin{equation}
\label{tau2mu}
\tau_2^{\mu}=p^{\mu}\;\tau_2-(-1)^{-j-l-m}\;T_2^{\mu}\,,
\end{equation}
and

\begin{equation}
\label{tau2munu}
\tau_2^{\mu\nu}=-p^{\mu}p^{\nu}\;\tau_2+p^{\mu}\;\tau_2^{\nu}+
p^{\nu}\;\tau_2^{\mu}+(-1)^{-j-l-m}\;T_2^{\mu\nu}\,.
\end{equation}

Particular cases for $T_1$, $T_2$ and $\tau_2$ such as $T_1(-1,-1,0)$,
$T_2(-1,-1,-1,0)$, etc., can be worked out from the general expressions. All
the above results are in agreement with the ones tabulated in
\cite{leib,bass,milgram}.

\section{Discussion and Conclusion.}

NDIM is a technique wherein the principle of analytic continuation plays a key
role. We solve a ``Feynman-like'' integral, i.e., a negative dimensional loop
integral with propagators raised to positive powers in the numerator and then
analytically continue the result to allow for negative values of those
exponents and positive dimension. 

In positive dimensions, Feynman integrals for covariant gauge choices can be
worked out with a variety of methods. However, when we work in the light-cone
gauge things become more complicated in virtue of the presence of unwieldy
gauge dependent singularities. And herein comes the help of NDIM with
surprising effect: propagators raised to positive powers in the
``Feynman-like'' integrals does not have poles of any kind to trouble us. {\em
Therefore no prescription is needed} in the NDIM approach, and moreover, {\em
no partial fractioning is necessary}. The beauty and the strength of NDIM to
deal with light-cone integrals is revealed and demonstrated in a marvelous way.

So, we can summarize all this by enumerating the outstanding features of NDIM:
i) No prescription at all is required to deal with gauge dependent poles of the
usual Feynman integrals; ii) The overall structure of the Feynman integrals in
the light-cone gauge is preserved, i.e., there is no need to introduce factors
of the form $q\!\cdot\!n^*$ in denominators as prescription input; iii) There
is no need to use parametrization of any kind, so that there are no parametric
integrals to solve; iv) There is no need to perform integration with split
components such as in \cite{leib-nyeo}, where the integration in space-time is
performed by decomposing $d^{2\omega}\!q \rightarrow d^{2\omega-1}{\bf
q}\,dq_4$; v) There is no need to resort to partial fractionings such as
(\ref{d-formula}); vi) The final result comes out for arbitrary {\it negative}
exponents of propagators, so that special cases of interest are all contained
therein; vii) The final result is already within the dimensional regularization
context.

In this work we calculated integrals --- scalar, vector, and second-rank
tensor --- pertaining to light-cone gauge with arbitrary exponents of
propagators and dimension. Our results given in (\ref{t1}), (\ref{t1mu}),
(\ref{t1munu}), (\ref{t2}), (\ref{t2mu}), (\ref{t2munu}), (\ref{tau2}),
(\ref{tau2mu}), and (\ref{tau2munu}) can be worked out for particular values
for the exponents and compared to those existing in the literature and checked
that they are all in agreement.

But, with no doubt, the most outstanding conclusion that we can draw from this
exercise is that {\em no prescription} was required to tackle the light-cone
singularities. Of course, it is a matter of straightforward generalization
that all other non-covariant gauge choices will follow suit.

\acknowledgments{\ A.G.M.S. gratefully acknowledges FAPESP (Funda\c c\~ao de
Amparo \`a Pesquisa do Estado de S\~ao Paulo, Brasil) for financial support. }

\end{document}